\documentclass[a4paper,11pt]{article}
\usepackage{jcappub}
\usepackage{siunitx}
\usepackage{graphicx}
\usepackage{xcolor}
\title{\boldmath A volcanic chronosequence as a time-resolved paleo-detector array to study the cosmic-ray flux in the Late Pleistocene and Holocene}

\author[a,1]{Claudio Galelli,\note{Corresponding author.}}
\author[a]{Lorenzo Caccianiga,}
\author[b,a]{Lorenzo Apollonio,}
\author[c,d]{Paolo Magnani,}
\author[e]{and Vincent Breton}

\affiliation[a]{INFN - Sezione di Milano, Via Celoria 16, 20133 Milan, Italy}
\affiliation[b]{Dipartimento di fisica "A. Pontremoli", Università degli Studi di Milano, Via Celoria 16, 20133 Milan, Italy}
\affiliation[c]{Gran Sasso Science Institute, Viale F. Crispi 7, 67100 L'Aquila, Italy}
\affiliation[d]{INFN - Sezione LNGS, 67100 Assergi, L'Aquila, Italy}
\affiliation[e]{Laboratoire de Physique de Clermont, 4 Av. Blaise Pascal, 63170 Aubière, France}

\emailAdd{claudio.galelli@mi.infn.it}

\abstract{We present a phenomenological study demonstrating the feasibility of using olivine xenoliths from the Chaîne des Puys as a time-resolved paleo-detector array to probe the cosmic-ray flux over the last 40,000 years. This volcanic region provides a unique chronosequence of samples brought to the surface by well-dated eruptions. By modeling the expected density of nuclear recoil tracks induced by cosmic-ray muons in olivine, we show that the signal is detectable and above backgrounds from natural radioactivity. We demonstrate that by analyzing samples with different exposure ages, it is possible to construct a time-differential measurement of the cosmic-ray flux. This method shows sensitivity to historical variations, such as the enhanced flux expected during the Laschamp geomagnetic excursion ($\sim$41~kyr) and the potential contribution from nearby supernovae, for which we use the Antlia supernova remnant precursor as a benchmark. This work establishes a new application of the paleo-detector technique for long-scale time-domain high-energy astrophysics and provides direct scientific motivation for experimental efforts to measure these track records.}


\arxivnumber{2510.23126}

\begin{document}

\maketitle
\flushbottom

\section{Introduction}

The emerging field of paleo-detectors aims to utilize natural minerals as long-exposure particle detectors, offering a unique opportunity to probe astrophysical phenomena and fundamental particle physics over geological timescales~\cite{Baum2023}. The concept underlying this field is the search for linear defects left by particles in the crystalline structure of suitable minerals, called paleo-detectors. This idea has been proposed in recent years, primarily as a technique to detect dark matter~\cite{Baum:2018tfw, Drukier:2018pdy} and neutrinos~\cite{Baum:2019fqm}. These studies take advantage of the enormous exposure that can be acquired through age, even with a small amount of material. 

The primary mechanism of interest for cosmic-ray studies using paleo-detectors is the interaction of high-energy secondary particles from atmospheric showers, predominantly muons, with the nuclei of the crystal lattice. While a muon passage itself does not leave a track, its interaction can induce spallation, causing the target nucleus or a fragment of it to recoil with significant kinetic energy. This recoiling nucleus, often a new cosmogenic isotope, travels a short distance through the lattice, leaving behind a damage track with a length ranging from hundreds of nanometers to tens of micrometers. The density of these accumulated tracks is proportional to the integrated particle flux, while the distribution of their lengths is correlated with the recoil energy spectrum of the produced isotopes. By measuring these features, one can, in principle, reconstruct the history of the cosmic-ray flux at Earth, providing a novel tool for probing past high-energy astrophysical events~\cite{Galelli:2025, Caccianiga2024}. 

Various experimental techniques have been proposed for the detection and characterization of tracks in paleo-detectors. Optical microscopy is the primary candidate for measuring muon-induced signals, as it allows for the efficient scanning of large sample areas to identify features in the micrometer to tens of micrometers range. The contrast of these tracks is typically enhanced through chemical or plasma-based etching. While highly effective for rapid data acquisition, optical methods are generally limited by their lower spatial resolution compared to more advanced nanoscopic techniques. Furthermore, the necessary sample preparation, including cutting and etching, and the two-dimensional measurements, can introduce geometric distortions into the observed track-length spectrum, a factor that must be accounted for in the final analysis.

Our previous work established the viability of this technique, proposing that evaporite minerals from the Messinian Salinity Crisis ($\sim$6~Myr ago) could serve as a detector for the muon flux of that epoch~\cite{Caccianiga2024}. We demonstrated that such a sample could be sensitive to variations in the primary cosmic-ray flux, such as those expected from nearby supernovae.

In this paper, we extend and evolve this methodology from a single-epoch measurement to a time-resolved analysis. We present a new phenomenological study focused on olivine xenoliths, inclusions from mantle minerals in lavic flows, from the Chaîne des Puys volcanic field in Auvergne, France. This region provides a unique geological scenario: a chronosequence of samples brought to the surface by a series of well-dated, distinct volcanic eruptions spanning the late Pleistocene and early Holocene ($\sim$40,000 to 7000 years ago)~\cite{Miallier2012, Boivin2017}. During eruptions, the host basaltic magma, with temperatures exceeding 1000°C, acts as a natural furnace, annealing any pre-existing tracks from the xenolith's previous geological history. This process ensures that the track record in any given sample begins precisely at the time of its host eruption. Subsequent eruptions in the volcanic chain source mantle material of similar characteristics, creating a series of well-defined, sequential, and overlapping exposure windows. We aim to establish the experimental feasibility of this approach and lay the groundwork for transforming the paleo-detector concept into a time-resolved observatory for probing the recent history of our galactic neighborhood.

\section{The Geological Scenario: Volcanism in the Chaîne des Puys}

\begin{figure}[ht!]
 \centering
 \includegraphics[width=0.4\columnwidth]{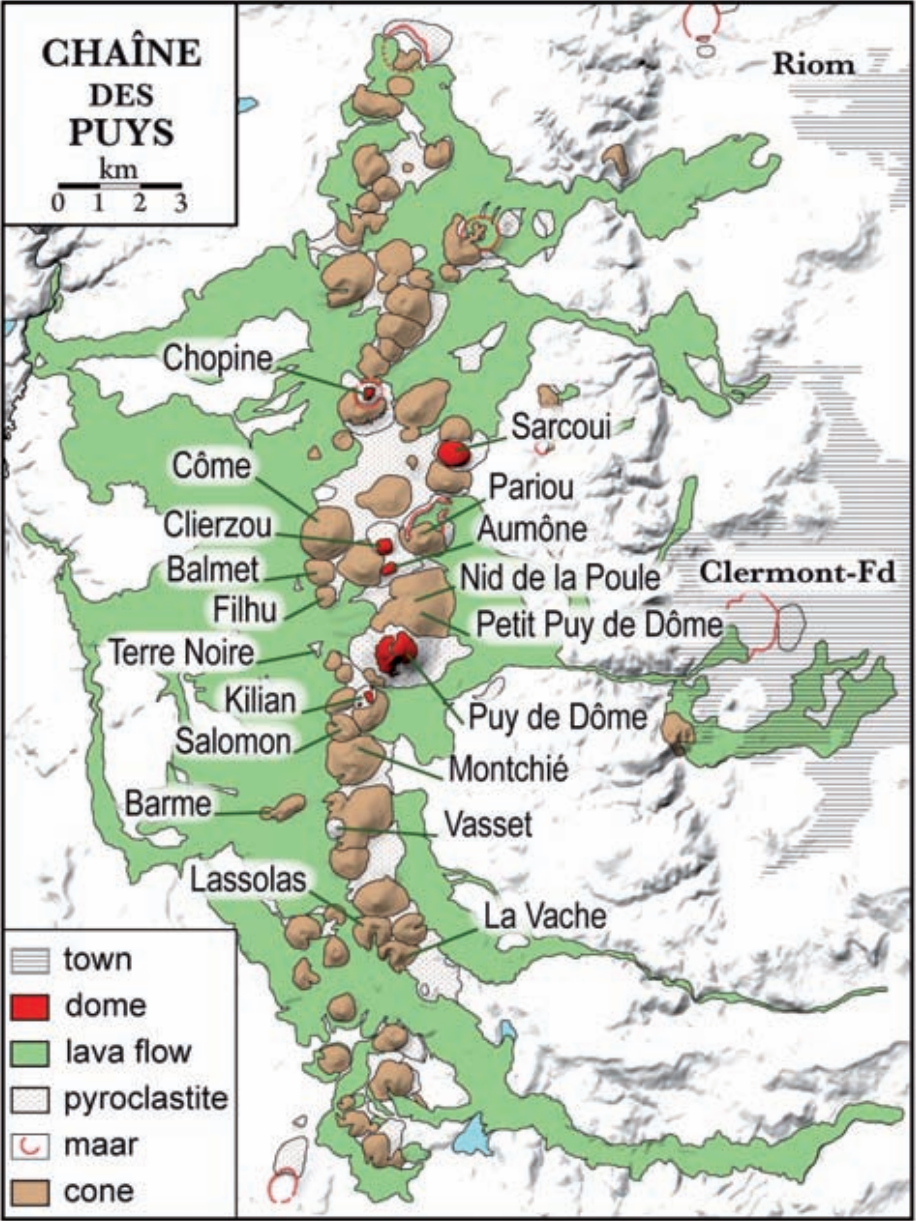}
 \caption{A map of the main part of the Chaîne des Puys volcanic field, west of Clermont-Ferrand.}
 \label{fig:map}
\end{figure}

The experimental foundation of this study rests upon the unique geological setting of the Chaîne des Puys volcanic field in the Massif Central mountain range, in the region of Auvergne, France (Figure~\ref{fig:map}). This region, a UNESCO World Heritage site, provides a natural laboratory ideally suited for our scope. The field consists of a dense, 40~km-long, north-south oriented alignment of approximately 80 monogenetic volcanoes, all formed within the last 95,000 years as a result of magma ascending through fractures associated with the Limagne rift system~\cite{Boivin2017}.

The volcanic activity is characterized by relatively small, discrete eruptions that transported mantle material to the surface. Of particular importance for our work are the numerous eruptions that carried olivine-rich peridotite xenoliths. Xenoliths are fragments of the Earth's upper mantle that are transported to the surface by an eruption, manifesting as inclusions in the solidified lava. The presence of these inclusions is a well-documented feature of many lava flows across the chain, providing a direct source of our target mineral~\cite{Boivin2017}.

The target mineral for this study, olivine ((Mg, Fe)$_2$SiO$_4$), was chosen for several key reasons that make it an excellent paleo-detector candidate. It is very common and geologically stable, ensuring its presence and the long-term preservation of tracks. Its abundance in mantle xenoliths, especially in peridotites, makes it readily available in volcanic scenarios worldwide, such as the one presented here. Furthermore, olivine has good optical transparency, which facilitates the analysis of internal tracks using microscopy techniques. These properties, together with the identification of extremely ancient olivines, have generated significant interest in the material within the broader paleo-detector community~\cite{Hirose:2025}, which is investigating its potential for a wide range of astroparticle physics searches.

The power of this geological scenario lies in its nature as a chronosequence, providing an array of independent paleo-detectors, each activated at a different point in time. By analyzing samples from this sequence, we can measure the total integrated track density over multiple, nested time windows. If the cosmic-ray flux has remained constant, the total number of tracks in each sample should be strictly directly proportional to its age. Therefore, any measured deviation from this linear relationship would serve as possible evidence for a period of increased or decreased cosmic-ray activity in the past, offering a powerful tool for probing the time-variability of our local astrophysical environment.

For this study, we have selected a representative series of well-dated eruptive events summarized in Table~\ref{tab:chrono} with the respective references. A particularly interesting target is represented by samples whose exposure window overlaps with the Laschamp geomagnetic excursion. This event, dated to approximately 41~kyr, was one of the last major excursions of Earth's magnetic field. During this period, the dipole field intensity dropped to as low as 10-25\% of its present-day value~\cite{Boivin2017, Laj2014}. This reduction in geomagnetic shielding would have allowed a greater flux of low-energy galactic cosmic-rays to penetrate the atmosphere~\cite{MUSCHELER2005}, enhancing the production of secondary particles. The Chaîne des Puys contains lava flows which erupted during this excursion, such as the Coulée de Laschamp, from which the magnetic event takes its name, and the Coulée d'Olby. Our choice of the Lemptégy II eruption ($\sim$30~kyr) also allows us to probe the period following the recovery of the magnetic field, providing a crucial point of comparison.

\begin{table}[htbp]
 \centering
 \caption{Chronology of selected volcanic events in the Chaîne des Puys, with corresponding literature reference.}
 \label{tab:chrono}
 \begin{tabular}{l c l}
 \hline
 \textbf{Volcanic Event} & \textbf{Age (kyr)} & \textbf{References} \\
 \hline
 Puy de Montcineyre & 7.65$\pm$0.12 & \cite{Juvigne2016, Boivin2017}\\
 Puy de la Vache & 8.64$\pm$0.06 & \cite{Miallier2012, Boivin2017}\\
 Puy Pariou & 9.5$\pm$0.5 & \cite{Miallier2012, Boivin2017}\\
 Puy de Dôme & 10.96$\pm$0.15 & \cite{Miallier2010, Boivin2017}\\
 Puy de Côme & 13.1$\pm$0.7 & \cite{Miallier2012, Boivin2017}\\
 Puy de Lemptégy II & 30$\pm$4.5 & \cite{Guerin1983, Boivin2017} \\
 Laschamp Event & 41.4$\pm$1.1 & \cite{Laj2014, Boivin2017}\\
 \hline
 \end{tabular}
\end{table}

\section{Modeling Track Production in Olivine}

To quantify the expected signal in the Chaîne des Puys xenoliths, we have developed a comprehensive simulation pipeline based on the framework established in our previous work~\cite{Caccianiga2024}. The full analysis code is publicly available\footnote{The complete analysis code and input files for this work can be found at \url{https://github.com/cgalelli/PrimusCode}.}. The goal of the pipeline is to translate a given cosmic-ray flux history and geological scenario into the observable number and length distribution of nuclear recoil tracks per unit mass.

The simulation chain is a three-step process. First, we model the primary cosmic-ray flux and its propagation through the atmosphere. This is handled by the \texttt{MCEq} (Matrix Cascade Equation) package~\cite{Fedynitch2015, Riehn:2017}, which calculates the resulting sea-level muon flux. The primary cosmic-ray spectrum input to \texttt{MCEq} is controlled by the \texttt{crflux} package \footnote{\texttt{https://crfluxmodels.readthedocs.io/en/latest/}}. To simulate historical variations, we have implemented a custom class inheriting from \texttt{crflux.PrimaryFlux}. This allows us to apply on-the-fly modifications to a baseline model, enabling the simulation of scenarios such as an enhanced flux from a nearby supernova or from a geomagnetic excursion like the Laschamp event. For this work, we simulate a "normal" scenario consistent with the currently measured cosmic-ray spectrum using \texttt{crflux.HillasGaisser2012}~\cite{Gaisser2012}, and an alternative scenario ("SN250") including an additional, now-extinct cosmic-ray component from a supernova with parameters consistent, as a possible realistic example, with the Antilia supernova: age $\sim$50~kyr, distance $\sim$250~pc~\cite{Fesen2021}.

Second, we model the muon interactions within the mineral using \texttt{Geant4}~\cite{Agostinelli2003}. We simulate 10$^4$ muons per energy bin, using 100 bins in logarithmic scale from 1 MeV, below which we do not anticipate enough recoil energy for track formation, to 10 TeV, where the muon flux is always strongly suppressed. The particles are propagated through a cylinder of olivine of radius 10 m and height 100 m. The simulation records the species and recoil energy of the nuclides produced by the interaction of the muons with the component nuclei of the mineral. A weight is then applied, in energy bins, depending on the flux obtained from \texttt{MCEq} and the estimated overburden depth on top of the sample, assuming a sample slice of 0.01 mm. 

Third, we compute tables of the projected range of each nuclide in the material as a function of kinetic energy using the SRIM (Stopping and Range of Ions in Matter) software~\cite{Ziegler2010}. These tables are convolved with the recoil energy obtained by the \texttt{Geant4} simulation to compute the rate dR/dx of track production as a function of track length.

A similar approach is taken for the two main track-generating backgrounds, radiogenic neutrons and fission fragments of Uranium. For both, the energy spectrum is taken from tabulated values and convoluted with SRIM tables for the relevant nuclide for fission fragments, and with a specific function in the \texttt{WIMPy-NREFT} python package~\cite{WIMpy-code} for radiogenic neutrons.

\begin{figure}[ht!]
 \centering
 \includegraphics[width=0.45\columnwidth]{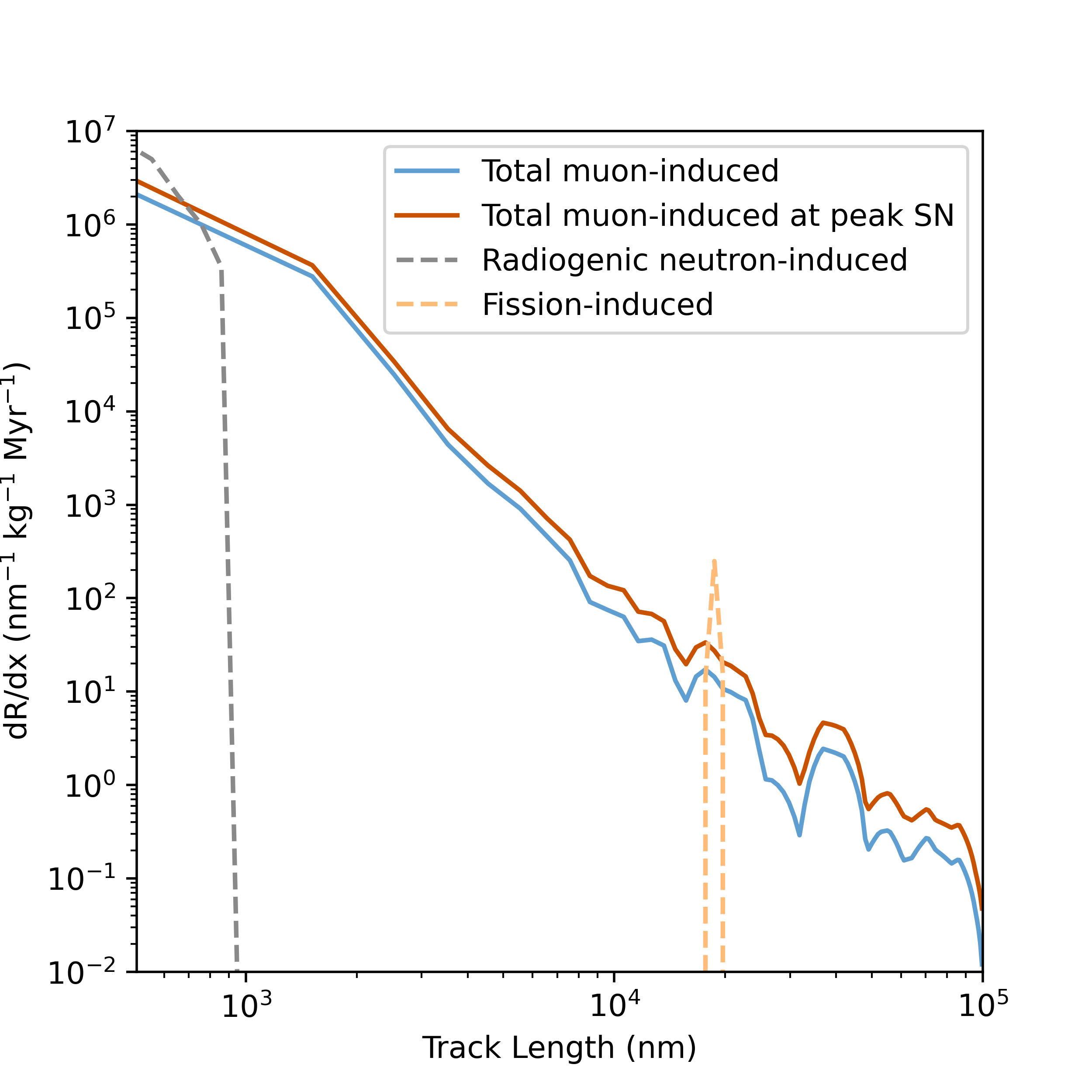}
 \includegraphics[width=0.45\columnwidth]{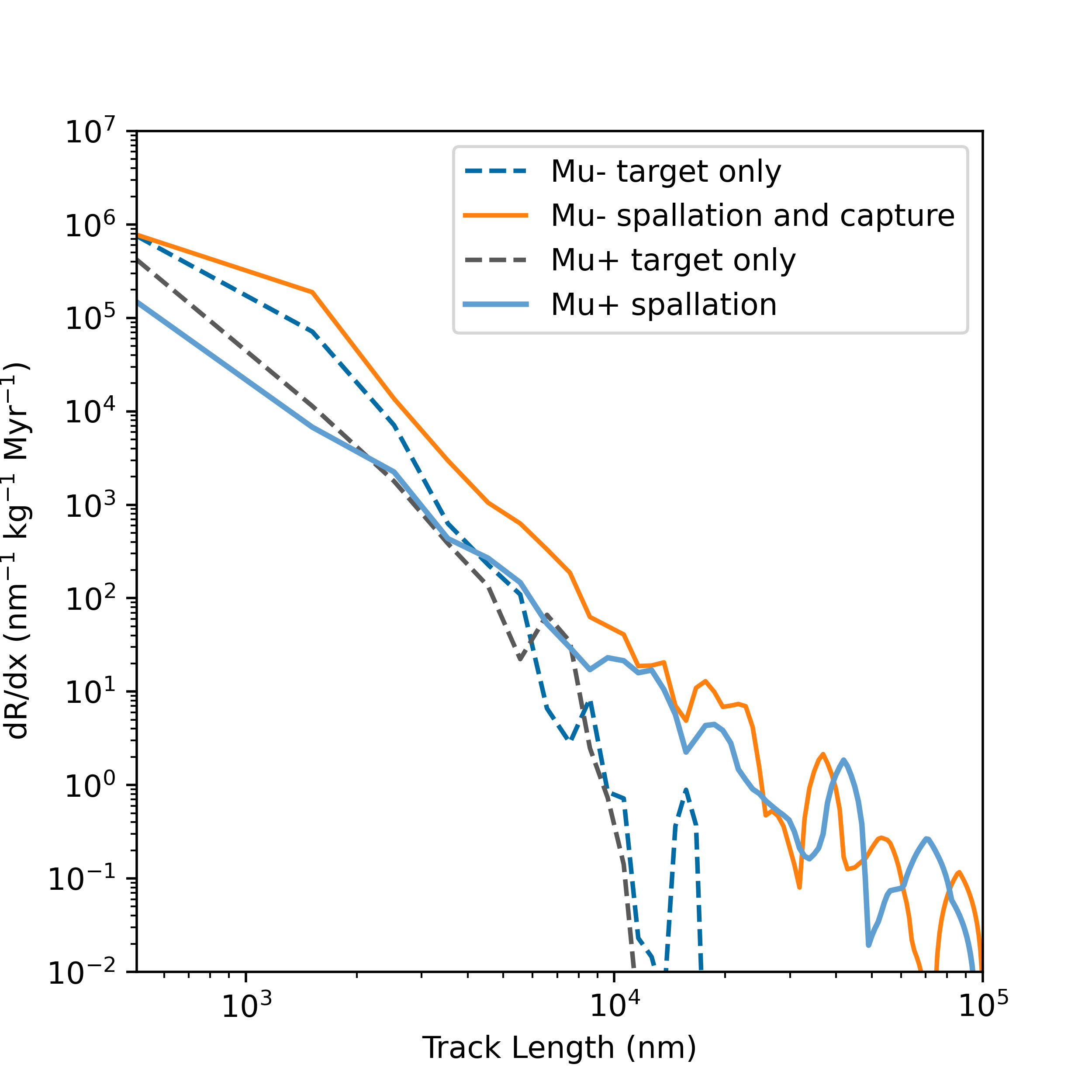}
 \caption{Differential track production rate in olivine. Left: Total muon-induced rate for the "normal" (blue) and "SN250" (brown) scenarios. Backgrounds from radiogenic neutrons (grey, dashed) and fission fragments (light orange, dashed) are shown. Right: Component breakdown of the muon signal. Dominant contributions from isotopes produced via $\mu^-$ spallation/capture (orange, solid) and $\mu^+$ spallation (orange, solid) are compared to subdominant elastic recoils from $\mu^-$ (blue, dashed) and $\mu^+$ (grey, dashed).}
 \label{fig:rate}
\end{figure}

\begin{figure}[hb!]
 \centering
 \includegraphics[width=0.7\columnwidth]{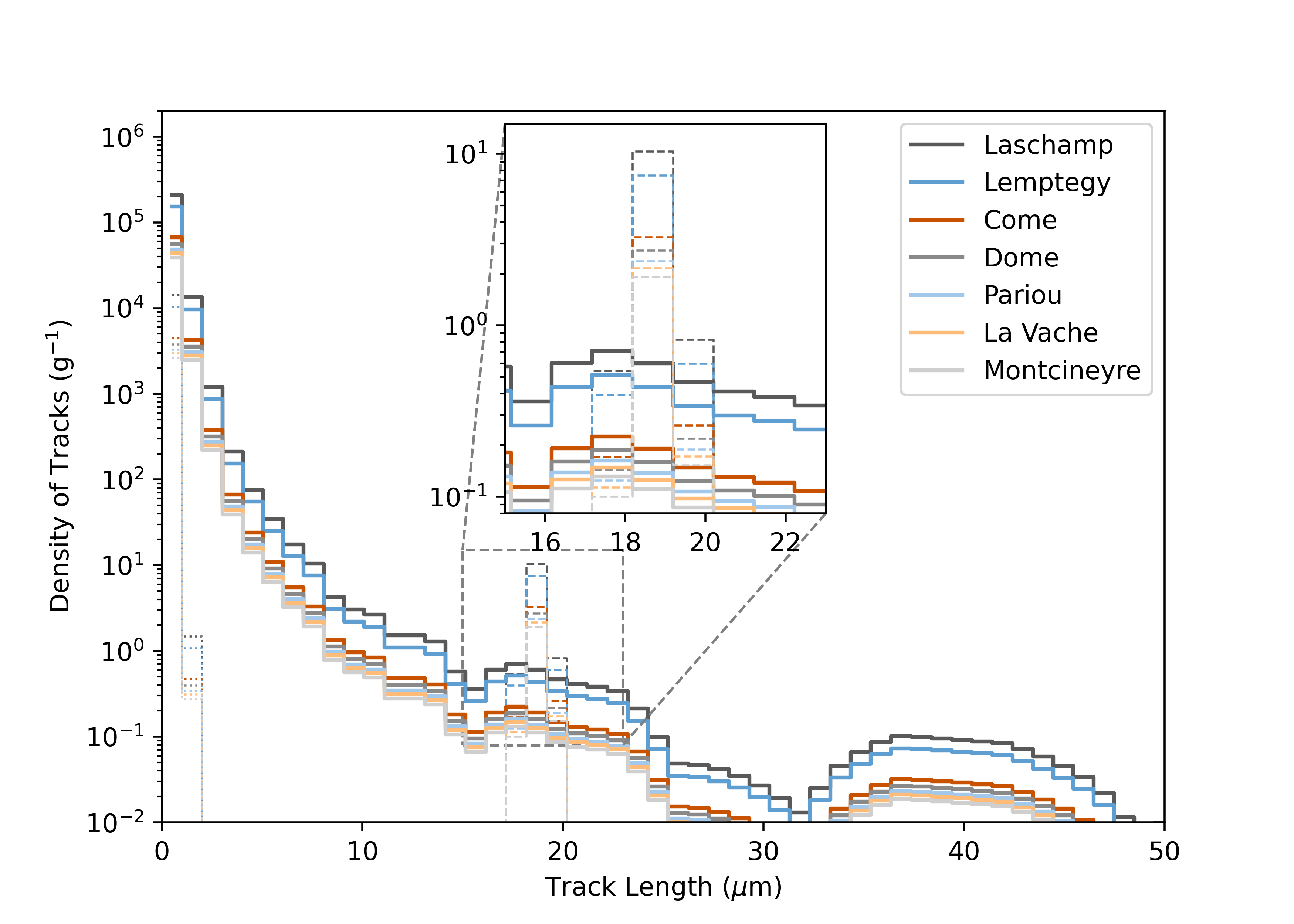}
 \caption{Predicted number of tracks per gram of olivine for each volcanic sample, in the normal flux scenario. The muon-induced track spectra are shown for all eruptions, scaling with their respective exposure times, while the background from fission and neutrons is shown in dashed lines.}
 \label{fig:number_per_gram}
\end{figure}

\section{Results and discussion}

The primary output of our simulation is the differential track production rate as a function of track length, presented in Figure~\ref{fig:rate}. We compare the muon-induced signal for both the "normal" and "SN250" scenarios against primary backgrounds: spontaneous fission of $^{238}$U and radiogenic neutrons. Assuming a 10~ppb Uranium concentration typical for mantle xenoliths, the cosmic muon signal exceeds the radiogenic neutron background for tracks longer than $\sim$\SI{2}{\micro \meter}. The fission track rate is comparable in magnitude to the muon signal in the relevant length interval. A decomposition of the muon-induced rate reveals that the spectrum is dominated by recoiling cosmogenic isotopes produced by negative muon spallation and capture. In contrast, elastic recoils of the constituent mineral nuclei are subdominant and confined to track lengths below \SI{10}{\micro \meter} for both $\mu^-$ and $\mu^+$.

Integrating this rate over the specific exposure time for each volcanic scenario yields the total number of tracks per gram of material, our primary observable. In figure~\ref{fig:number_per_gram}, we show these predicted track length spectra for all selected eruptions only in the normal scenario. The spectra for the muon signal clearly scale with their respective exposure times, while the fission track background is shown in overlapping dashed lines. This plot also highlights that the expected signal always remains above the background level in the relevant track length region.

\begin{figure}[ht!]
 \centering
 \includegraphics[width=0.7\columnwidth]{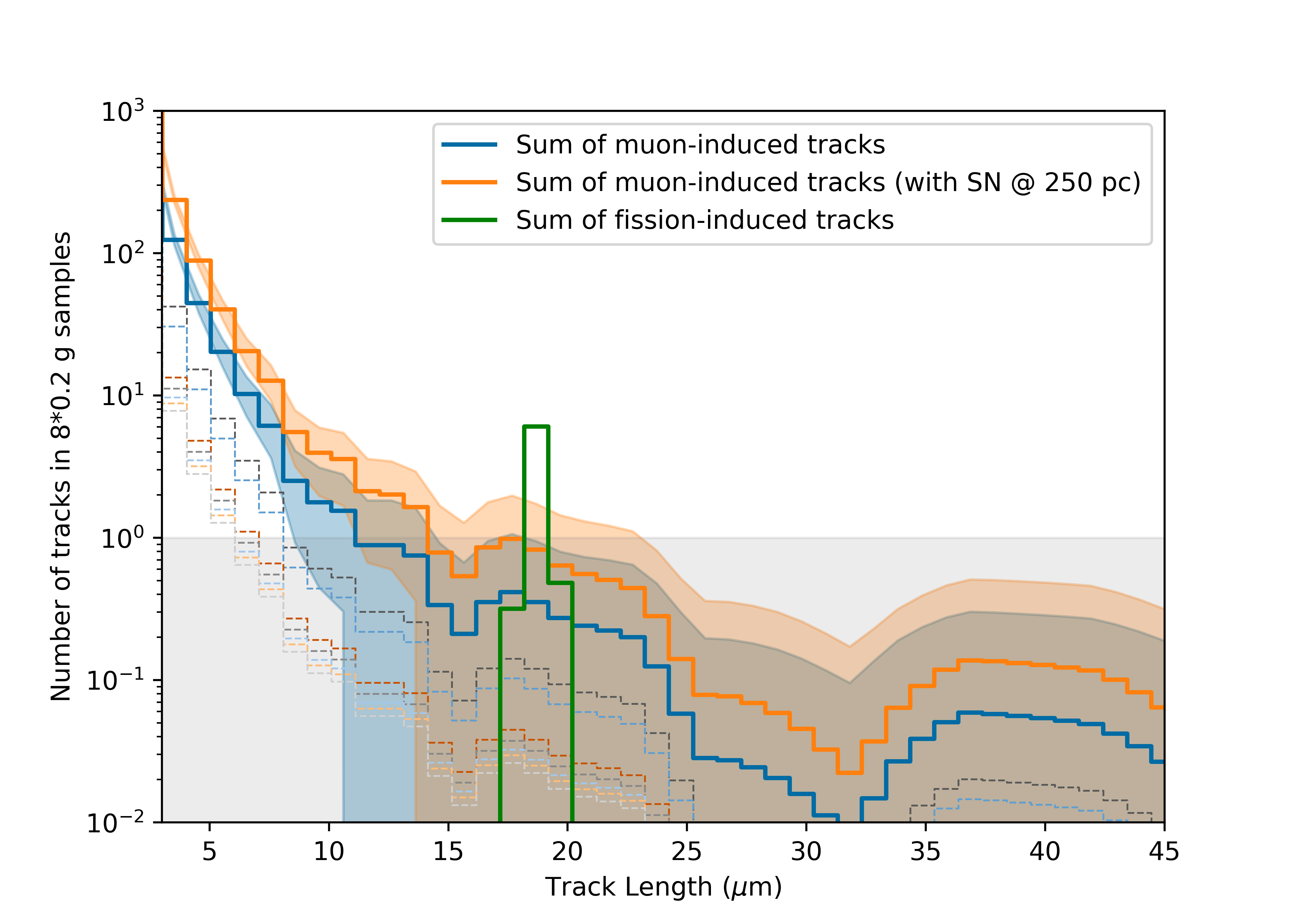}
 \caption{Total number of muon-induced tracks expected from a collection of 0.2-gram samples, one from each volcanic eruption. The cumulative signal is shown for both the normal cosmic-ray flux scenario (blue) and the SN250 (orange), with bands representing a Poissonian uncertainty. The dashed lines indicate the contribution from each individual eruption in the normal case. The green line represents the expected summed counts of all the fission tracks.}
 \label{fig:total_collection}
\end{figure}

\begin{figure}[ht!]
 \centering
 \includegraphics[width=0.7\columnwidth]{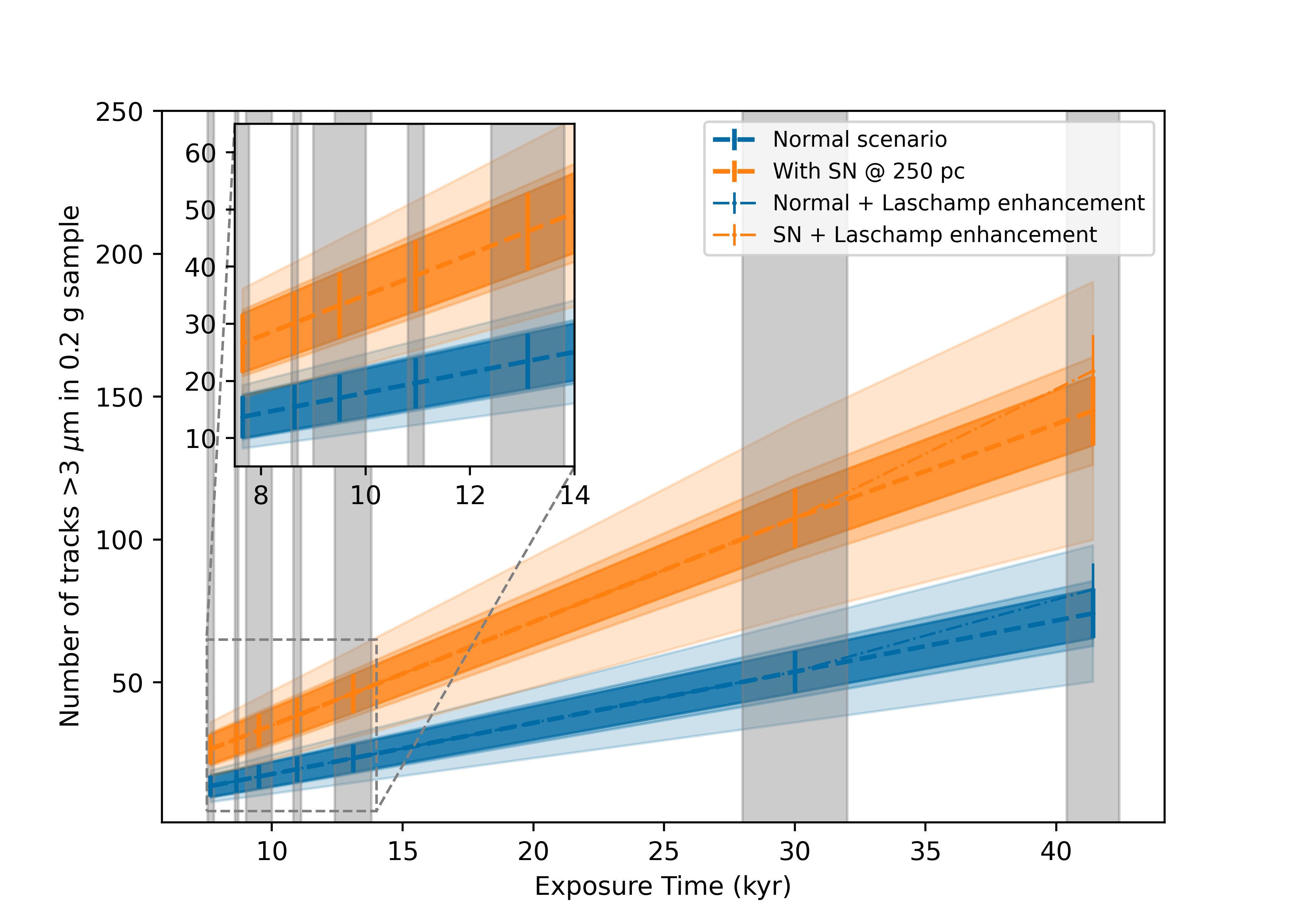}
 \caption{Time evolution of the total integrated number of muon-induced tracks as a function of sample exposure time. The points represent the expected signal for each of the volcanic scenarios in the normal (blue) and SN250 (orange) flux scenarios, with Poissonian, Poissonian + 10\% counting error, and Poissonian + 30\% counting error bands. For each flux, an enhancement due to the lower Earth's magnetic field during the Laschamp event is shown as a dashed line deviating from each flux scenario in the time interval between 30 and 40~kya. The inset shows a zoom in the 7 to 14~kyr range, an interval of time in which eruptions were particularly frequent. Time uncertainties in the eruption dating are represented by the grey vertical bands.}
 \label{fig:time_evolution}
\end{figure}

For an experimental campaign, the total number of tracks that can be measured is a crucial figure of merit. Assuming an analysis of 0.2 grams of olivine from each site, we predicted the total signal expected from the entire sample collection. This cumulative track spectrum for both the normal and the SN250 flux scenarios with a Poissonian $\sqrt{N}$ uncertainty band is shown in figure~\ref{fig:total_collection}. This illustrates that separation between the two flux scenarios is possible at least at the 1$\sigma$ level up to $\sim$\SI{15}{\micro \meter}, assuming complete analysis of a set of 0.2 g samples.

The ultimate goal of this method is to probe the time-variability of the cosmic-ray flux. In figure~\ref{fig:time_evolution}, we plot the integrated track number as a function of the exposure time for each volcanic scenario. As expected, the track count scales linearly with exposure duration, with slight deviations due to time integration steps of the astrophysical scenarios. The blue and dotted lines represent the expected track accumulation for the normal and SN250 flux models, respectively. The two scenarios are discriminated at least at the 1$\sigma$ level even when accounting for a possible 30\% counting systematic. For the predicted enhancement for samples exposed during the Laschamp event, modeled by an additional primary flux component in the GeV to 10 GeV range, the signal difference lands close to the 1$\sigma$ confidence interval even without accounting for systematics, and as such, should be hard to detect with this method. While more advanced statistical techniques could further refine our ability to distinguish between cosmic-ray flux models, their implementation must be carefully calibrated taking into account the chosen experimental setup. Instrumental effects such as finite spatial resolution or etching-induced biases, in the case of optical microscopy, can distort the observed track spectrum. As such, a more complete assessment of the discriminating power of the samples requires a comprehensive treatment of these experimental systematic uncertainties.

\section{Conclusions}

We have presented a phenomenological study demonstrating the feasibility of using olivine xenoliths from the Chaîne des Puys volcanic field as a time-resolved paleo-detector array. The unique geological history of this region provides a sequence of samples with well-defined exposure windows. Our simulation pipeline, which models the cosmic-ray flux, muon interactions in olivine, and the resulting nuclear recoil tracks, predicts a clear and detectable signal well above the expected backgrounds from natural radioactivity.

This work also shows that the chronosequence of eruptions allows for a differential measurement, enabling a direct probe of the flux's time-variability during the Late Pleistocene and Holocene. This technique shows possible sensitivity to specific phenomena, such as the potential contribution from nearby, now-extinct sources like the Antlia supernova, and in more optimistic cases, to weaker phenomena like the enhanced flux expected during the Laschamp geomagnetic excursion.

This provides a direct scientific antecedent for the PRImuS (Paleo-astroparticles Reconstructed with the Interactions of MUons in Stone) project~\cite{Galelli:2025}, an experimental effort funded by the INFN. The primary goal of PRImuS is to realize the measurements proposed in this paper and in our previous work on Halite. To this end, we have already begun the crucial first step of sample recovery, starting with olivine xenoliths sourced directly from the 7,650-year-old Puy de Montcineyre maar. The successful identification of muon-induced tracks in these and other samples from the Chaîne des Puys will mark a significant advance, transforming the paleo-detector concept into a new observational tool for time-domain high-energy astrophysics.

\acknowledgments
PRImuS is an INFN experiment funded by the CSN5 Young Scientist Grant 2024. The authors thank Didier Miallier, Thierry Pilleyre, Pierre-Jean Gauthier, Valentin Niess, Denis Andrault, and Emmanuel Gardes for the fruitful discussions in Clermont-Ferrand on xenoliths and olivine samples from the Chaîne des Puys and possible methods of analysis and readout. We also thank the members of the paleo-detector community, especially Alexey Elykov and Emilie LaVoie-Ingram, for their interest and exchange of ideas. 

\bibliographystyle{JHEP}
\bibliography{bibliography}

\end{document}